\documentclass[conference]{IEEEtran}

\usepackage{amssymb}
\usepackage{amsmath}
\usepackage{graphicx}
\usepackage{epstopdf}
\begin{document}
%
\title{Routing Diverse Crowds in Emergency with Dynamic Grouping}

\author{Olumide J Akinwande and Huibo Bi\\
Intelligent Systems and Networks Group\\
Dept. of Electrical and Electronic Engineering, Imperial College, London SW7 2AZ, UK\\
Email: \{olumide.akinwande13, huibo.bi12\}@imperial.ac.uk}

\maketitle

\begin{abstract}
Evacuee routing algorithms in emergency typically adopt one single criterion to compute desired paths and ignore the specific requirements of users caused by different physical strength, mobility and level of resistance to hazard. In this paper, we present a quality of service (QoS) driven multi-path routing algorithm to provide diverse paths for different categories of evacuees. This algorithm borrows the concept of Cognitive Packet Network (CPN), which is a flexible protocol that can rapidly solve optimal solution for any user-defined goal function. Spatial information regarding the location and spread of hazards is taken into consideration to avoid that evacuees be directed towards hazardous zones. Furthermore, since previous emergency navigation algorithms are normally insensitive to sudden changes in the hazard environment such as abrupt congestion or injury of civilians, evacuees are dynamically assigned to several groups to adapt their course of action with regard to their on-going physical condition and environments. Simulation results indicate that the proposed algorithm which is sensitive to the needs of evacuees produces better results than the use of a single metric. Simulations also show that the use of dynamic grouping to adjust the evacuees' category and routing algorithms with regard for their on-going health conditions and mobility, can achieve higher survival rates.
\end{abstract}


\begin{IEEEkeywords}
Emergency navigation, QoS driven protocol, Dynamic grouping, Cognitive Packet Network.
\end{IEEEkeywords}

%
\IEEEpeerreviewmaketitle

\section{Introduction}

The advancement in information technology (IT) has facilitated the emergence of complex emergency management systems (EMS) \cite{gelenbe2012emergency} based on in-situ wireless sensor networks (WSN). However, in most of the applications, routing algorithms use a single criterion to choose paths for all evacuees without considering their specific requirements due to variance on age, mobility, level of resistance to hazard etc. For instance, evacuees such as sick people or aged people need to choose the safest paths that will be well ahead of the spreading hazard while for others may prefer the quickest paths without hazard. In this paper, we propose a multi-path routing algorithm to tailor different services to diverse categories of evacuees. Moreover, the category that an evacuee belongs to can shift in terms of their on-going physical conditions. We use the concepts of the Cognitive Packet Networks (CPN) to search escape paths for the evacuees in a building based on predefined metrics. The remainder of this paper is organised as follows. In the next section we review the literature relevant to our work. Section \ref{CPN} presents the concept of CPN and its variations for the evacuee routing problem. The routing metrics used in CPN as well as dynamic grouping mechanisms are then introduced in Section \ref{routingmetric} and Section \ref{dynamicgrouping}, respectively. The simulation models and assumptions are described in Section \ref{model} and the experimental results are presented in Section \ref{results}. Finally, we draw conclusions in Section \ref{conclusion}.

\section{Related Work}
\label{related}

Disaster management and building evacuation can improve significantly with the help of IT solutions.  Initial research in this field were actually driven by defence applications \cite{Autonomous98}
including enhanced reality simulators \cite{Kaptan} and evacuation models that incorporated models of human mobility and behaviour \cite{gwynne1999review}. Recent survey articles \cite{kuligowski2005review,GelenbeW12} can assist in selecting  research directions with agent-based models that offer some level of realism by representing each individual evacuee as an agent that follows specific goals.

Research then moved further to the development of complex Emergency Cyber-Physical-Human systems to direct evacuees to exits in a real time \cite{filippoupolitis2009distributed} with sensor nodes (SNs)  responsible for collecting hazard information while the decision subsystem composed of decision nodes (DNs) provides advice to evacuees through visual indicators or portable devices. Li et al. \cite{li2003distributed} implement a WSN consisting of sensors that continuously monitor the environment and distribute a danger-level map across the network. Optimization methods have been suggested \cite{DBLP:journals/fi/Han13}
using distributed decision making with random neural networks \cite{DBLP:journals/neco/GelenbeT08,DBLP:journals/eor/Gelenbe00} to overcome the huge complexity of decision making for a large number of agents
in the presence of spatial information to help select exit routes and decide about the appropriate allocation of rescuers and technical assets.

Communications are essential in this context, but they can easily malfunction during emergencies. To ease this problem, in \cite{gelenbe2012wireless} a resilient emergency support system (ESS) disseminates emergency messages to evacuees with the aid of opportunistic communications (Oppcomms). Experimental results indicate that this system is quite resilient to network failures during an emergency. Because Oppcomms are susceptible to malicious attacks \cite{gelenbe2012emergency} a defense mechanism that uses a combination of identity-based signatures (IBS) and content-based message verification to detect malicious nodes is proposed, while in \cite{gelenbebi2014Emergency} an infrastructure-less emergency navigation system guides evacuees with the aid of smart handsets and cloud servers.
As the core of EMS, navigation algorithms also have motivated considerable research. Potential-maintenance approaches such as \cite{barnes2007emergency,tseng2006wireless} establish global navigation graphs via assigning each sensor a potential value which is determined by the distance to exits and/or the hazard area. Evacuees are directed towards the exit along the paths with the smallest cumulative potential. To avoid civilians being directed to hazards \cite{li2003distributed} one can use artificial potential fields where the exit generates an attractive potential which ``pulls'' the evacuees towards it while the sensors in dangerous areas generate a repulsive potential that ``pushes'' them away. However, these approaches require massive information exchanges to synchronise the navigation map of each sensor.

The notion of ``effective length'' \cite{filippoupolitis2009distributed} calculated as the product of the physical length and the hazard intensity, together with Dijkstra's algorithm can compute the best path to exits \cite{filippoupolitis2009emergency,filippoupolitis2009decision}, and can include \cite{filippoupolitis2012emergency}  information about the spatial hazard. The ``Uniformity principle''  \cite{francis1981uniformity} is also useful in showing that proper allocation of evacuees to routes, such that all exit routes have the same clearance time, is essential in minimizing evacuation time \cite{pursals2009optimal}, while in \cite{chalmet1982network,lu2005capacity} network flow models mimic evacuation planning problems and convert the original network to time-expended networks. To reduce the high computational cost caused of these linear programming methods, in \cite{desmet2014capacity} the Cognitive Packet Network is used for route discovery.

\section{Cognitive Packet Network}
\label{CPN}

\subsection{CPN Architecture}

The Cognitive Packet Network \cite{gelenbe1999cognitive,gelenbe2004} was initially proposed for packet routing in large scale packet networks, and was first suggested for evacuee routing in \cite{filippoupolitis2010emergencyPHD} via integrating with a $m$-sensible routing algorithm \cite{gelenbe2003sensible}. Unlike traditional packet network protocols where routers have all the intelligence, CPN constructs intelligence into packets for routing and flow control through a decentralised self-adaptive decision architecture \cite{gelenbe2009steps} offering a framework to implement learning algorithms and adaptation.

CPN contains three types of packets - smart packets (SP), acknowledgement packets (ACK) and dumb packets (DP). Each CPN node maintains a Mailbox (MB) storing diverse classes of QoS information grouped under path and associated QoS measurements, which is regularly updated by ACKs that traverse the node. A MB will discard expired QoS information or the worst one when reaches its capacity. SPs are sent by CPN nodes to explore the network and gather relevant information with respect to a user-specified QoS. The measurements made by the SPs represent their ``experience'' of the network state and future SPs exploit this in making better decisions. The preferred learning method is the RNN \cite{gelenbe1990stability} with reinforcement learning (RL) which penalizes or rewards SP decisions so that subsequent decisions can provide better results in meeting QoS goals. The QoS goals (routing metrics) which are detailed in Section \ref{routingmetric} are the inputs of RNN. When an SP reaches one exit, an ACK, carrying the SP's measurements, is generated by the destination node and travels back to source node along the discovered loop-free path. The ACK updates the MB of every node along its path and triggers the nodes to run the learning algorithms and update the relevant decisions. The DPs, which carry the payload, are always source-routed using the highest ranked path information. The DPs can also be used to carry out measurements. In summary, the first set of SPs sent aim to primarily establish connection between the source and the intended destination while the subsequent SPs update the path to optimising a given QoS metric. To improve performance and avoid unnecessary burden of the network, lost packets will be discarded. An SP is considered lost if it has traversed a set number of hops or for a set time without reaching its destination while an ACK or a DP is lost if enters a node that is not along its specified path.

\subsection{Variations for Evacuee Routing Problem}

CPN is suitable for resource-limited emergency environments because each CPN node can adaptively collect information with ``interested'' counterparts rather than synchronise data with all the other nodes in the environment. To adapt the original CPN to emergency evacuation problems, we assume that a wireless sensor network, which consists of sensor nodes (SNs) and decision nodes (DNs), is deployed in the built environment. Each DN is considered as a CPN node and can emit SPs to other DNs to explore the environment. Guiding evacuees in the presence of a spreading hazard is similar to routing packets in a fast-changing packet network. By using CPN, each source node in a computer network can discover routes to destinations with respect to the pre-defined QoS goals and DPs can then follow the top ranked routes to achieve certain QoS requirements. Similarly, during an emergency, each decision node in the EMS sends SPs to search physical paths to exits for the evacuees in proximity. Therefore, evacuees can be considered as DPs. However, unlike DPs, evacuees cannot be source-routed as they need path updates in case the initial path becomes dangerous or congested. This represents a major challenge as, ideally, an evacuee should only get an updated path if the present path becomes bad, but since only a path's source node will know if an old path becomes bad and by that time, an evacuee already given the old path would have left and therefore cannot be informed. Based on our assumption that all evacuees can receive full path information, we use a movement depth value that ensures every evacuee traverse a given number of nodes before getting an update. The optimal movement depth value for any emergency scenario, will be a function of the evacuee distribution and possibly the decision algorithm. In our treatment, the movement depth of evacuees is set to 3.

\section{Routing Metrics}
\label{routingmetric}

\begin{figure}[!ht]
\centering
\includegraphics[width=0.26\textwidth]{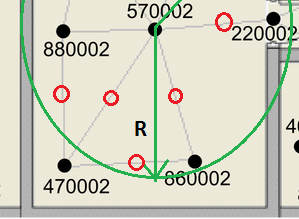}
\caption{DNs are located on the black dots while SNs are positioned on the red rings. SNs in the green circle belong to the spatial set of $N_{570002}$.}
\label{fig: spatialsetEEE}
\end{figure}

The routing metrics are the QoS goals that are pursued by SPs and optimised by the RNN algorithm. When an ACK brings back sensory data to the source node, collected information will be extracted and then measured by routing metrics, the result will be used as input of RNN. In this section, we propose two metrics that integrate with spatial information gathered by SNs to guide evacuees. As shown in Figure \ref{fig: spatialsetEEE}, we hypothesize that DNs are located at the vertices of the building graph and calculate advices for evacuees in proximity. SNs are located along each edge and provide real-time information with respect to the conditions of their immediate environment that will be relevant in computing the optimal paths. Spatial information of a DN $i$ is obtained by virtue of observations from a set of nearby SNs, which is defined as $N_i$. A SN $s$ is considered to belong to set $N_i$ if the Euclidean distance between the SN $s$ and the DN $i$ is not greater than a certain distance $R$, i.e., $ N_i := \lbrace s \in S : d(i,s)\leq R \rbrace $ where $S$ is the set of all SNs in the building and $d(i,s)$ represents the Euclidean distance between a DN $i$ and a SN $s$. Each SN estimates the hazard intensity $H$ along the edge it is located as:
\begin{equation*}
H = \left\{
\begin{array}{ll}
1 & \textrm{if no hazard is present} \\
k \cdot 10^3  & \mbox{otherwise}
\end{array}\right.
\end{equation*}
where $k$ is an integer in the interval $[1,8]$ that indicates the level of the detected hazard. Each SN also records the time instant when it first detects the hazard.

Each DN can obtain the hazard intensity estimated by the SNs along all its incident edges. The DNs can also obtain sensory data from all SNs within their spatial set when $R$ has a nonzero value. A DN $i$ uses these measurements to compute the effective length of all its incident edges as follows:
\begin{equation}
\label{fireintensity}
E_{\pi(s)} = l_{\pi(s)} \cdot [H_{\pi(s)} + \frac{\sum_{\substack{j = 1; j \neq s}}^n H_{\pi(j)} }{n-1}]
\end{equation}
where $\pi(s)$ represents the edge that SN $s$ locates on. $E_{\pi(s)}$ is the effective length of the edge $\pi(s)$; $l_{\pi(s)}$ is the physical length of the edge $\pi(s)$; $H_{\pi(s)}$ is the estimated hazard intensity by sensor $s$; term $n$ is the number of sensors belonging to the spatial set $N_i$ of the DN. If $R=0$, the summation in (\ref{fireintensity}) evaluates to zero and the effective length becomes:
\begin{equation}
E_{\pi(s)} = l_{\pi(s)} \cdot H_{\pi(s)}
\end{equation}
When the hazard is first detected, each DN computes its initial predicted hazard reach time as:
\begin{equation}
t_{initial} = \frac{d(i,f)}{H_s}
\end{equation}
where $d(i,f)$ depicts the Euclidean distance between a DN $i$ and the fire source location $f$, $H_s$ is the estimated hazard spread rate. Similarly, when $R$ is nonzero and one of the SNs in a DN’s spatial set first detects the hazard, the DN can also compute the hazard reach time from the SN as:
\begin{equation}
\label{equ: predn}
t_{predn} = \frac{d(i,s)}{H_s}
\end{equation}
where $d(i,s)$ represents the Euclidean distance between a DN $i$ and the hazardous sensor location $s$, $H_s$ is the estimated hazard spread rate.

A DN regularly computes its hazard time $t_{haz}$, which is the estimated time before a node becomes hazardous, as follows:
\\* $\circ$ If $R \neq 0$ (considering spatial information)
\begin{equation}
t_{haz} = \min\{(t_{initial}- t_{elapsed}), \min_{s \in N}[t_{predn}^{s}- t_{enode}^{s}]\}
\end{equation}	
\\* $\circ$ Otherwise
\begin{equation}
t_{haz} = t_{initial}- t_{elapsed}
\end{equation}
where $t_{elapsed}$ is the time since the hazard started; $t_{predn}^s$ is the initial predicted time for the hazard at SN $s$, which belongs to spatial set of the DN, to reach the DN, evaluated using \ref{equ: predn}; $t_{enode}^s$ is the time elapsed since the sensor $s$ detected the hazard. If a SN in the spatial set of a DN has not detected the hazard, the DN sets its value for $(t_{predn}^s-t_{enode}^s)$ to a very large number.		

The two metrics that we consider in the experiments are described as follows. Time metric customises paths for prime-aged adults (``class one group'') while safety metric is associated with children and the elderly (``class two group'').

\textbf{Time Metric:} the primary goal of this metric is to choose egress paths that minimise the evacuation times of the evacuees. The path traversal time is estimated as follows:
\begin{equation}
\label{equ: time}
G_t = \sum_{\substack{i =1}}^{n-1}[\frac{E(\pi(i,i+1))}{V} + t_{delay}]
\end{equation}
where $\pi$ represents a particular path and $n$ is the number of DNs on it. Term $i$ depicts the $i$-th DN on the path. $E(\pi(i,i+1))$ is the effective length of the edge $\pi(i,i+1)$ connecting the nodes $i$ and $i+1$ and $V$ is the speed of the evacuee.

We assume an evacuee at a node to be a customer waiting to be served by a DN where the service is obtaining the best escape path based on the operating metric. We also assume that only one evacuee can be served at a time by a DN so that when evacuees arrive at a DN while it is busy, the evacuees will be considered to be in a queue. Therefore, $t_{delay}$ estimates the possible congestion delays that might be experienced by an evacuee at the nodes of an escape path using the Little's theorem as follows:
\begin{equation}
\label{equ: delay}
t_{delay} =\frac{ q_i + t_i \cdot (a_i - d_i)}{a_i}
\end{equation}
where $q_i$ is the present queue length of node $i$, $t_i$ is the time for an evacuee to reach node $i$ and $a_i$ and $d_i$ are the average arrival and average departure rates of node $i$. The DNs are used to estimate these values. Term $t_{delay}$ is set to zero if (\ref{equ: delay}) evaluates to a negative value so as to comply with the non-decreasing property of a QoS metric stated in \cite{gelenbe2003sensible}.

\textbf{Safety Metric:} in this case, we seek paths such that the evacuees are ahead of the spreading hazard. For a path between any node $i$ and a destination or exit node $n$, this metric is evaluated by the hazard intensity of a route:
\begin{equation}
\label{equ: safety}
G_s = \sum_{\substack{i =1}}^{n-1}[L [t_i -t_{haz}^i] \cdot c_i + s(\pi(i,i+1))]
\end{equation}
where $t_i$ is the predicted time for an evacuee to reach node $i$, $t_{haz}^i$ is the computed hazard time of node $i$, $c_i$ is the hazard growth rate at node $i$ and $s(\pi(i,i+1))$ is a safety value for the edge between nodes $i$ and $i+1$. The safety value ensures that $G_s$ never evaluates to zero for an entirely safe path which prevents the inverse of the metric ($G_s^{-1})$) from being undefined. In our treatment, we set this safety value as the effective length of the edge. The function $L[x]$ takes the value $x$ if $x$ is nonnegative and zero if $x$ is negative.

For the time metric, we include the spatial information in order to prevent the system from directing the evacuees close to hazardous regions, while for the safety metric, the spatial information helps in updating the initial prediction of the hazard time. The level of the spatial information used by the DN is determined by the value of $R$.

In our simulations, each DN stores information relevant to evaluating the path metrics given in (\ref{equ: time}) and (\ref{equ: safety}). Therefore, a DN should compute the following values:

\begin{itemize}
  \item The effective length $H$ of all its incident edges.
  \item Its hazard time $t_{haz}$.
  \item The estimates of its queue length, average arrival and average departure rates.
\end{itemize}

An estimate of the average arrival rate is regularly updated by the DN using the number of the evacuees that have arrived the node during a certain time period. The average departure rate is set to $1$ as this is the average time cost for a DN to serve an evacuee. To minimise the complexity of our problem, we choose a constant value of $20 cm/s$ as the value of the hazard spread rate for the simulations.

\section{Dynamic Grouping Mechanisms}
\label{dynamicgrouping}

To the best of our knowledge, previous studies in emergency navigation persist on a single decision algorithm during the whole evacuation process and do not adjust in accordance with individuals' physical conditions and their immediate environments. Although we have divided evacuees into two groups with separate routing metrics in Section \ref{routingmetric}, we believe it is necessary for evacuees to be able to switch groups during an evacuation if certain conditions are fulfilled. For instance, when an individual which belongs to the ``class one group'' gets injured, it should be adapted to the safer ``class two group'' due to the reduced mobility and injury. Hence, in this section, we present two dynamic grouping mechanisms to enable evacuees to change groups and the associated algorithms rather than stick to the pre-defined groups.

This first mechanism is very simple: any evacuee in the first category whose health level has dropped below 50\% of its original value will be immediately considered as a class two agent. Hence, instead of choosing paths with the shortest time to exits, evacuees will choose safest escape paths. The other mechanism is proposed based on the observation that certain vertices located in broad areas such as a hall can still have a long queue and the connected paths are not sufficiently used. These vertices are normally linked with ``bottlenecks'' such as staircases where stampede or continuous congestion may occur. Although both the time metric and safety metric take predicted congestion level of a route into consideration, the effect of other factors in the metrics such as ``effective length'' may induce evacuees at a vertex to choose an identical path. To exclude the influence of other factors, we consider the potential number of congestion encountered on a path as a secondary QoS metric and propose a congestion-alleviate mechanism to balance the main QoS requirement and the secondary QoS need. This congestion-ease policy makes use of the mailbox of CPN nodes and chooses a less congested path with acceptable main QoS value (safety level or egress time) rather than the top ranked path when the congestion level is high. More specifically, if the queue length at a vertex is larger than or equals to a certain value (In our treatment, we set it to 4), the related DN will re-assign the newly arrived evacuee to the congestion-ease group and suggest a less congested path.

\section{Simulation Model and Assumptions}
\label{model}

\begin{figure}[!ht]
\centering
\includegraphics[width=0.4\textwidth]{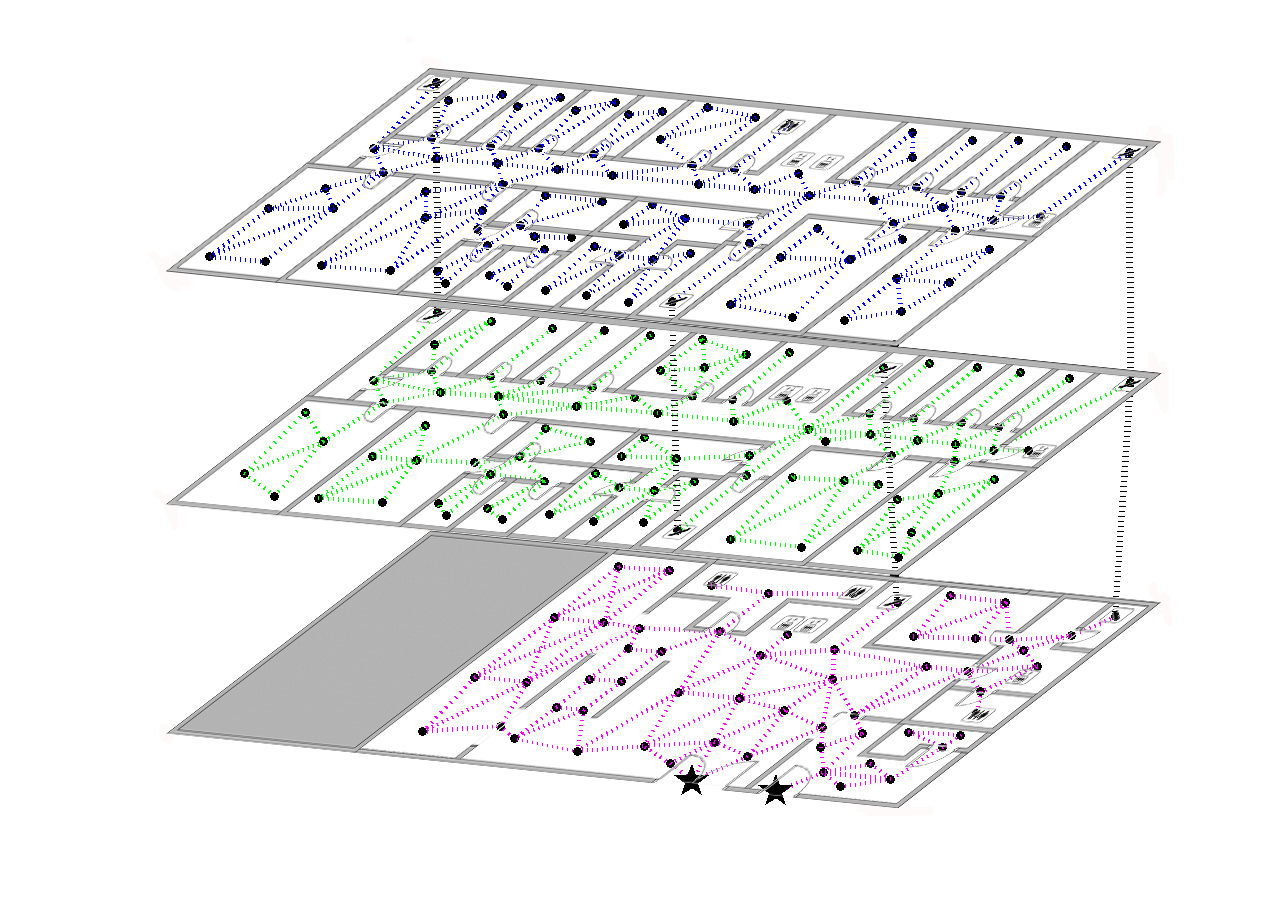}
\caption{Graph representation of the building model. Vertices (black round dots) represent locations where people can congregate such as rooms, doorways and corridors while the two black stars on the first floor depict exits.}
\label{fig: gmodelEEE}
\end{figure}

Experiments are conducted on a Java-based multi-agent simulation tool, namely Distributed Building Evacuation Simulator (DBES) \cite{filippoupolitis2009distributed,dimakis2010distributed}, which can import various building models to mimic confined spaces and use autonomic actors to imitate crowd behaviours. The building model in this paper simulates the three lower floors of the EEE building at Imperial College London as shown in Fig. \ref{fig: gmodelEEE}.

Evacuees are divided into ``class one agent'' (adults, 16 - 50 years) and ``class two agent'' (children or aged people) in accordance with their age. The two categories adopt different mobile and health models in which class one agent has higher mobility and higher resistance to fatigue and the hazard. The initial health level of simulated evacuees is set to 100 and will decrease due to fatigue and exposure to the hazard. If the health level of an evacuee falls below $50$, its speeds will also drop to half of the initial value. Initially, evacuees are randomly scattered in the building and are assumed to carry portable devices that can receive advices from the emergency management system.

\section{Experiments and Results}
\label{results}

To evaluate the performance of the proposed algorithm, we simulate a fire-related disaster in the building model with diverse occupancy rates (30, 60, 90 and 120 evacuees). The proposed algorithm which combines time metric and safety metric is examined under different levels of spatial information. The level of spatial information is determined by the operating communication range ($R$) of DNs. For comparison purposes, two algorithms which use time or safety metric individually are also experimented.

\begin{figure}[!ht]
\centering
\includegraphics[width=0.5\textwidth]{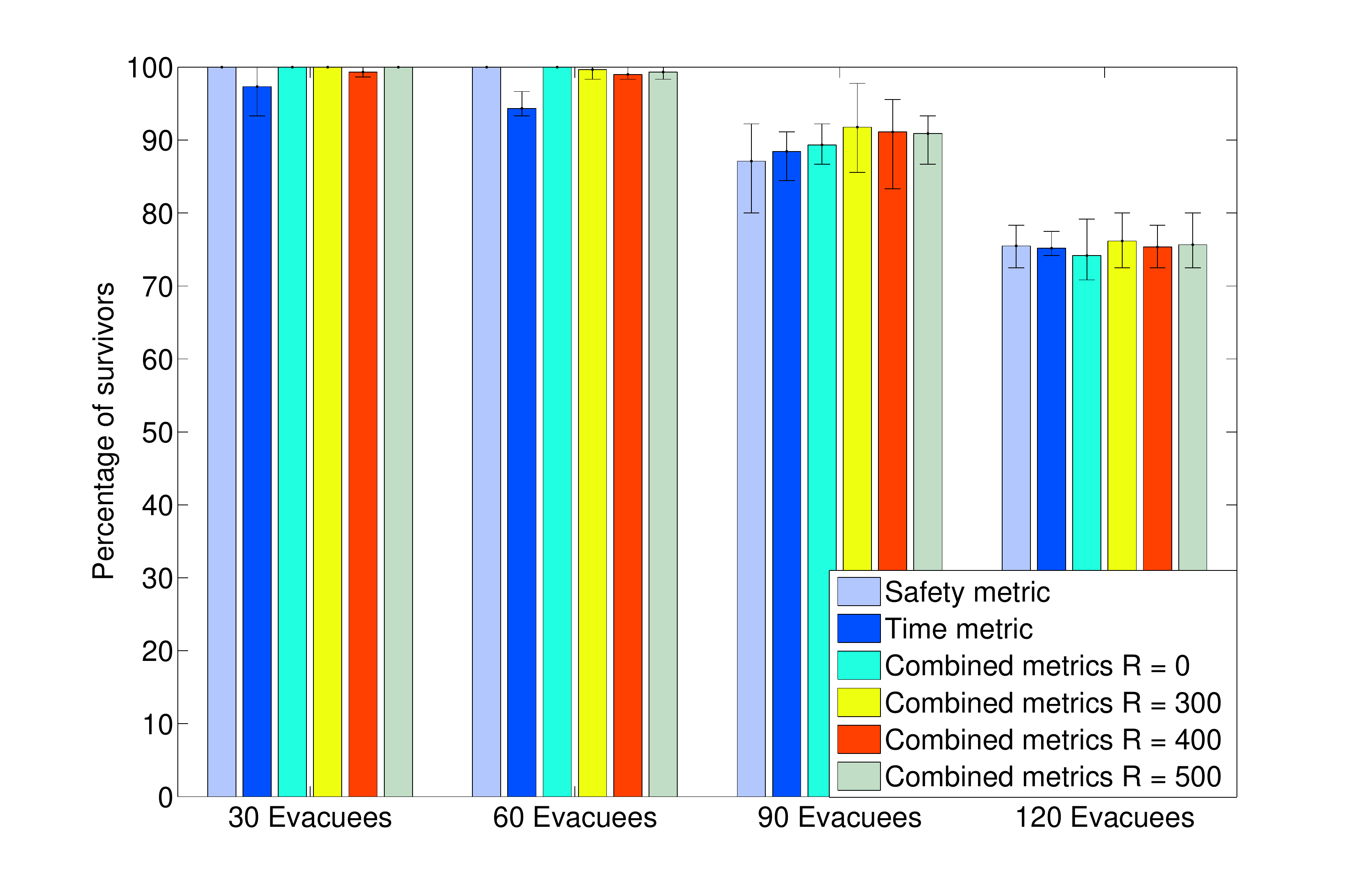}
\caption{The average survival rates for the decision algorithms. The experimental results are the average of 5 simulations, the error bars show the min/max result in any of the five simulation runs.}
\label{fig: survivalEEE}
\end{figure}

\subsection{Average Survival Rate}

As shown in Figure \ref{fig: survivalEEE}, for low levels of occupancy (30 and 60 evacuees), the safety metric (SM) achieves best performances on its own, while the time metric (TM) gives the worst performance overall. This is because unlike SM which tends to guide all the evacuees to the safest path, evacuees using TM may take the risk to traverse potential hazard areas in order to reduce the evacuation time. However, some evacuees may perish due to the decreased mobility caused by injury. The combined metrics (CMs) reach the performance of SM because they can naturally ease congestion by generating separate channels for two categories of evacuees. Moreover, rather than stick to the quickest path with higher risk, injured evacuees can switch to the safest path due to the benefit of dynamic grouping and therefore reduce fatalities.

On the other hand, for high population densities (90 and 120 evacuees), CMs perform the best among all the algorithms. This is because the congestion level has a considerable impact on the system performance in densely-populated scenarios. For 90 evacuees, because CMs can assign evacuees to a third group when severe congestion occurs (queue length $ > 3$) and suggest a less congested path with an acceptable QoS level, paths in broad areas are more sufficiently used and evacuees can reach exits with less latency. In addition, at this level, TM can be seen to perform better than SM. This is because TM can alleviate congestion more efficiently in comparison with SM. On the other hand, for 120 evacuees, CMs only perform slightly better than other two metrics, the reason is that the congestion-ease mechanisms are invalid to some extent in this scenario due to the extremely high occupancy.

In summary, CMs gain better results than other algorithms and CM with $R=300$ achieves the overall best performance. This reflects that the range of spatial hazard information ($R$) can affect the system performance. When $R$ is too small, evacuees are insensitive to the hazard and may choose risky routes. Hence, spreading hazard may block the chosen route and evacuees will have to traverse a detour path. On the contrary, if $R$ is too large, some paths with acceptable safety level may also be excluded and this can induce insufficient use of routes. Towards comparison algorithms, SM gains good performance in low occupancy rates but bad performance in high occupancy rates. This is because in order to keep all the evacuees far away from the hazard, SM only directs evacuees to safest paths and has the potential to cause jamming. When the population is low, this issue is negligible but can induce continuous congestion in densely-populated scenarios. Conversely, TM achieves unacceptable performance in low occupancy rates but improved results in high occupancy rates. This is because the quickest paths tend to traverse areas with potential risks and evacuees with low health level may perish owing to the fast spreading of the hazard as well as significantly reduced mobility. In high population densities, the embedded congestion ease mechanism of TM is less effective because injured evacuees with significantly reduced mobility become ``obstacles'' for other civilians and induce continuous congestion.

\subsection{The Effect of Dynamic Grouping}

\begin{figure}[!ht]
\centering
\includegraphics[width=0.5\textwidth]{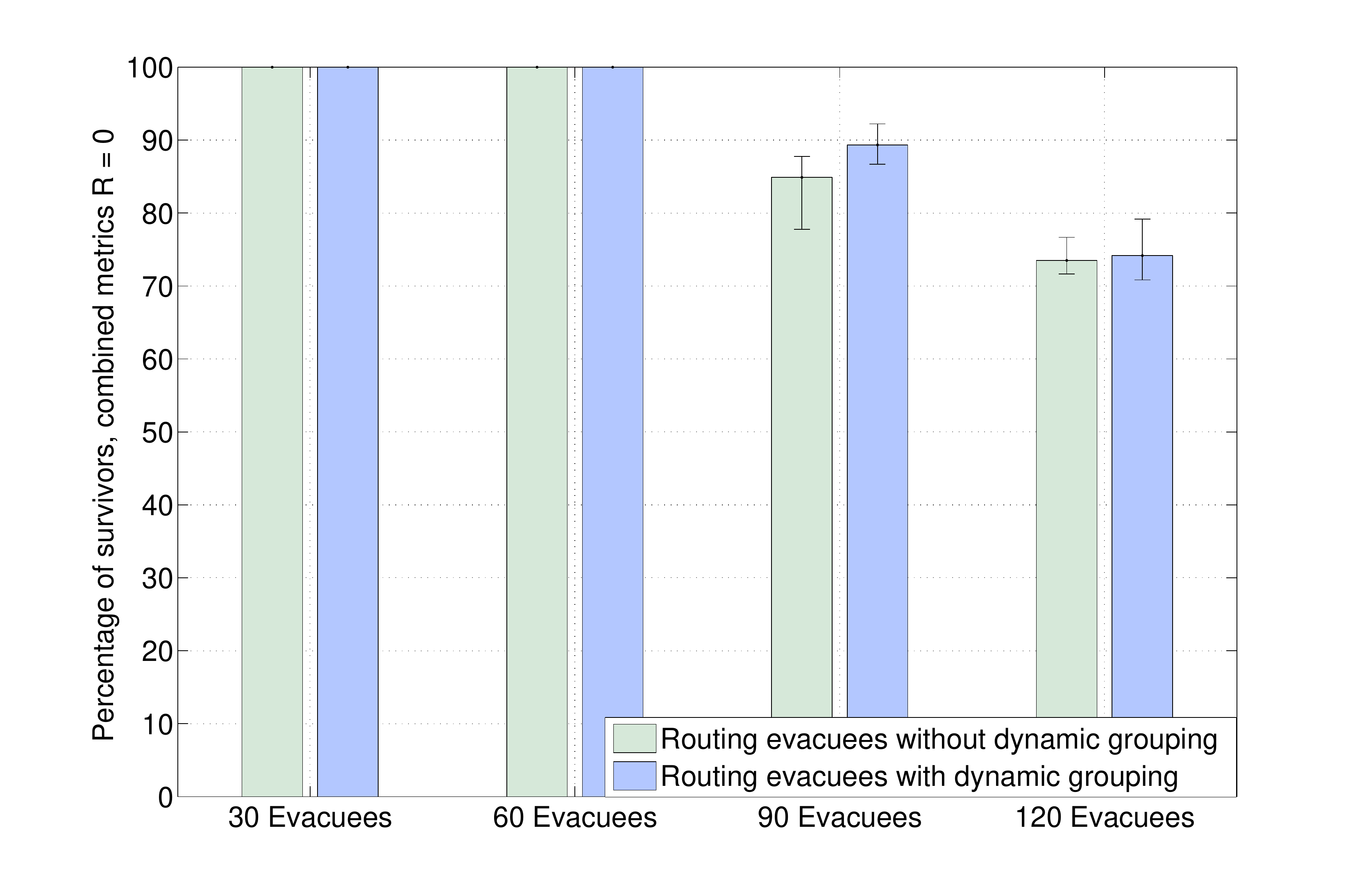}
\caption{Percentage of survivors, Combined metrics with $R=0$. The results are the average of 5 simulations, the error bars show the min/max value in any of the five simulation runs.}
\label{fig: survival2EEE}
\end{figure}

\begin{figure}[!ht]
\centering
\includegraphics[width=0.5\textwidth]{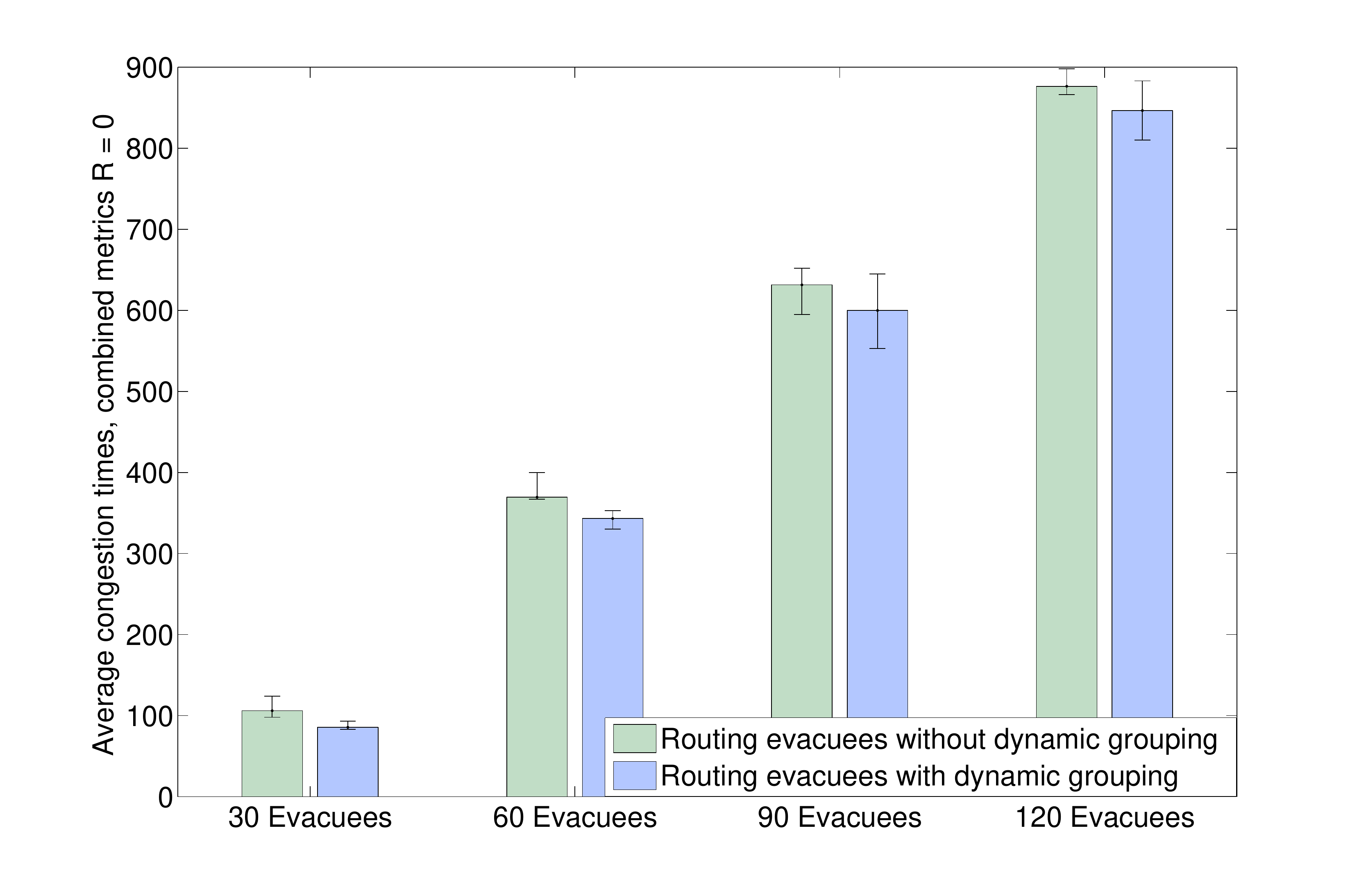}
\caption{Average congestion times, Combined metric with $R=0$. The results are the average of 5 simulations, the error bars show the min/max value in any of the five simulations.}
\label{fig: congestion2EEE}
\end{figure}

In the simulations where we consider CMs, a class one evacuee whose health level falls below $50$ will be immediately considered as a member of second class. Moreover, an evacuee that reaches a vertex with more than 3 individuals will be assigned to the congestion-ease group and choose a less popular path. In order to evaluate the effect of the ``dynamic grouping'', we repeat the experiments by routing two categories of evacuees with SM and TM separately and avoid changing of class. This means that class one agents will use TM throughout each simulation. Figures \ref{fig: survival2EEE} and \ref{fig: congestion2EEE} show the comparisons of results with $R = 0$.

The results indicate that the dynamic changing of groups generally has a positive impact on the system performance. Figure \ref{fig: survival2EEE} shows that dynamic grouping can improve the survival rate in highly population densities. From Figure \ref{fig: congestion2EEE} we can see that the use of congestion ease mechanism introduced in Section \ref{dynamicgrouping} can reduce the congestion level of evacuee flows.

\section{Conclusion and Future work}
\label{conclusion}

In this paper, we propose a multi-path routing algorithm to direct different types of evacuees with respect to their on-going requirements. Spatial hazard information is combined into routing metrics of CPN to prevent evacuees from being guided into hazards and provide a more accurate prediction on the fire spreading rate. Two dynamic grouping mechanisms which adjust the type and the associated decision algorithm of evacuees are presented with regard to evacuees' physical conditions and surrounding environments. The results indicated that this QoS driven routing algorithm provides improved performance and the use of dynamic grouping can achieve higher survival rates. The simulation results also imply that appropriate setting of parameters such as the range of spatial hazard information can significantly improve the performance of a routing algorithm. Hence, future research will focus on cloud-based fast-than-real-time simulation to seek optimal parameters or choosing appropriate navigation algorithms from initial conditions, and look at how search \cite{PhysRev}, communications \cite{Multihop} and energy consumption \cite{Energy} can be optimised.

\bibliographystyle{IEEEtran}
\bibliography{olumidemanagingcrowds}

\end{document}